\documentclass[aps,prd,12point,twocolumn,nofootinbib,showpacs,superscriptaddress]{revtex4-1}

\usepackage{amssymb}
\usepackage{graphicx}
\usepackage{amsmath, amsthm}
\usepackage{epstopdf}
\usepackage{hyperref}

\usepackage{amsmath,amsfonts,amssymb,mathrsfs}
\usepackage{graphicx}
\usepackage{color}
\def\be{\begin{equation}}
\def\ee{\end{equation}}
\begin{document}
\title{Revisiting the analogue of the Jebsen-Birkhoff theorem in 
Brans-Dicke gravity}

\author{Valerio Faraoni} \email{vfaraoni@ubishops.ca} 
\affiliation{Department of Physics \& Astronomy and STAR 
Research Cluster, Bishop's University, 2600 College St., 
Sherbrooke, Qu\'ebec, Canada J1M~1Z7} 

\author{Fay\c{c}al Hammad} \email{fhammad@ubishops.ca} 
\affiliation{Department of Physics \& Astronomy and STAR 
Research Cluster, Bishop's University, 2600 College St.,
Sherbrooke, Qu\'ebec, Canada J1M~1Z7} 
\affiliation{Physics Department, Champlain 
College-Lennoxville, 2580 College Street, Sherbrooke,  
Qu\'ebec, Canada J1M~0C8}

\author{Adriana M. Cardini} 
\email{a.marina.cardini@gmail.com} 
\affiliation{Department of Physics \& Astronomy, Bishop's University, 2600 
College Street, Sherbrooke, Qu\'ebec, Canada J1M~1Z7}

\author{Thomas Gobeil} 
\email{Thomas.Gobeil@usherbrooke.ca} 
\affiliation{Department of Physics, Universit\'e de Sherbrooke,  2500  
Blvd. de l'Universit\'e, Sherbrooke, Qu\'ebec, Canada J1K~2R1}
\affiliation{Department of Physics \& Astronomy, Bishop's University, 2600 
College Street, Sherbrooke, Qu\'ebec, Canada J1M~1Z7}

\begin{abstract}

We report the explicit form of the general static, spherically symmetric, 
and asymptotically flat solution of vacuum Brans-Dicke gravity in the 
Jordan frame, assuming that the Brans-Dicke scalar field has no 
singularities or zeros (except possibly for a central singularity). This 
general solution is conformal to the Fisher-Wyman geometry of Einstein 
theory and its nature depends on a scalar charge parameter. Apart from the 
Schwarzschild black hole, only wormhole throats and central naked 
singularities are possible.

\end{abstract}



\maketitle

\section{Introduction}
\label{sec:1}

In general relativity (GR), there is a unique spherical and asymptotically 
flat solution of the vacuum Einstein equations (with zero cosmological 
constant): the static Schwarzschild geometry. This fact, known as the 
Jebsen-Birkhoff theorem \cite{Birkhoff}, generalized in Ref.~\cite{Bronnikov1} to higher-dimensional GR, is extremely important because it 
makes GR black holes simple, there is no need to single out the 
``physical'' black hole solutions of GR, and the end point of 
gravitational 
collapse is completely determined. The Jebsen-Birkhoff theorem breaks down  
in 
theories of gravity alternative to GR, which are motivated by the need to 
explain the present acceleration of the universe without an {\em ad hoc} 
dark energy, and by unavoidable corrections to GR arising from any attempt 
to quantize gravity.  The prototypical alternative to GR is Brans-Dicke 
gravity \cite{BD}, which adds to the metric tensor $g_{ab}$ a scalar 
degree of freedom $\phi$, approximately corresponding to the inverse of 
the effective gravitational coupling strength, which becomes dynamical 
\cite{BD}. Brans-Dicke theory was generalized by 
promoting the constant Brans-Dicke parameter to a function of $\phi$ 
and/or   
by including a potential $V(\phi)$ \cite{ST}. More modern versions of 
scalar-tensor gravity include galileons, generalized galileons, and 
Horndeski theory, which are the subject of intensive research 
\cite{galileons}. The vacuum Brans-Dicke action\footnote{We use 
units in which 
the speed of light and Newton's constant are unity, and the notation of 
Ref.~\cite{Wald}.}
\be 
S_{BD}=\int d^4x \, \frac{\sqrt{-g}}{16\pi}  \left( \phi {\cal R} 
-\frac{\omega}{\phi} \, \nabla^c \phi \nabla_c \phi \right) \label{action}
\ee
(where ${\cal R}$ is the Ricci scalar and $\omega$ is the constant 
Brans-Dicke coupling) gives rise to the field equations 
\begin{eqnarray}
{\cal R}_{ab}-\frac{1}{2}\, g_{ab} {\cal R} &=& \frac{\omega}{\phi^2} 
\left( 
\nabla_a \phi \nabla_b \phi -\frac{1}{2} \, g_{ab} \nabla^c\phi 
\nabla_c\phi  \right) \nonumber\\
&&\nonumber\\  
&\, & +\frac{1}{\phi} 
\left( \nabla_a\nabla_b \phi  -g_{ab}\Box \phi \right) \,, 
\label{field1}\\
 &&\nonumber\\
\Box\phi &=& 0 \,. \label{field2}
\end{eqnarray}
Black holes in scalar-tensor gravity are not arbitrary: an important 
no-hair theorem due to Hawking  
states that all vacuum, stationary, and asymptotically flat black holes of 
Brans-Dicke gravity must reduce to Kerr black holes \cite{Hawking}. 
Hawking's no-hair theorem has been generalized to include more general 
scalar-tensor theories and a potential $V(\phi)$, provided 
that the latter has a minimum which allows for $\phi$ to sit in a state 
of equilibrium \cite{Bekenstein, SotiriouFaraoniPRL, BhattaRomanoPRL}. An  
essential feature in the proof of no-hair theorems is that $\phi$ 
becomes constant outside the horizon, reducing the theory to GR and any 
black hole to Kerr. A few maverick solutions are known 
which evade the no-hair theorems, but only at the price of 
physical pathologies such as the divergence of $\phi$ on the 
horizon (the solution of \cite{maverick} is one example, but it is 
linearly unstable \cite{unstable}). No-hair theorems for Horndeski and 
galileon theories, and ways to evade them, are the subject of a large  
literature (\cite{circumvent} and references therein).

In Brans-Dicke theory, things become more tricky. In fact, a 
theorem 
due to Agnese and La Camera \cite{ALC} states that in Brans-Dicke 
gravity the possible solutions describe only wormholes or naked 
singularities. 
Its 
proof 
is incorrect, as shown below. In any case, it is clear 
that this theorem contradicts Hawking's no-hair result and its 
generalizations \cite{Hawking, Bekenstein, SotiriouFaraoniPRL, 
BhattaRomanoPRL, JFnohair} because it excludes the Schwarzschild black 
hole which is indeed a solution, as is easy to verify. This Letter 
aims at presenting a comprehensive report that clarifies the issue by linking it to the most 
general solution of vacuum Brans-Dicke theory \cite{Bronnikov2}, 
which constitutes an analogue of the Jebsen-Birkhoff theorem of GR in 
Brans-Dicke theory, reworked here under the following physically 
reasonable assumptions:

\begin{enumerate}

\item the vacuum Brans-Dicke equations in the Jordan frame hold with 
$\omega\neq -3/2$;

\item the spacetime metric is spherically symmetric, static, and 
asymptotically flat (staticity  reflects a state of equilibrium, while 
asymptotic flatness characterizes isolated objects);

\item the Brans-Dicke scalar $\phi$ depends only on the radial coordinate 
$r$, it does not have poles or zeros (except possibly for a central 
singularity), and  $\phi(r) $ becomes constant as $r\rightarrow +\infty$.

\end{enumerate}

\section{The general Jordan frame solution} 
\label{sec:2}

Let us investigate the general solution under the assumptions above.

\subsection{The Agnese-La Camera theorem}

The Agnese-La Camera theorem \cite{ALC} states that, under the 
assumptions~1)-3), the only possible solutions describe wormholes or naked 
singularities. The proof of this theorem begins by writing the line 
element and scalar field as
\begin{eqnarray}
ds^2_{ALC} &=& -\left( 1-\frac{2\eta}{r}\right)^A dt^2 + \left( 
1-\frac{2\eta}{r}\right)^B dr^2  \nonumber\\
&&\nonumber\\
&\, & + \left( 1-\frac{2\eta}{r}\right)^{1+B} r^2 d\Omega_{(2)}^2 \,, 
\label{ALC1}\\
&&\nonumber\\
\phi_{ALC}(r) &=& \phi_0 \left( 
1-\frac{2\eta}{r}\right)^{\frac{-(A+B)}{2}} \,,\label{ALC2}
\end{eqnarray}
with $d\Omega_{(2)}^2=d\theta^2 +\sin^2\theta \, d\varphi^2$ and 
\be
1-\frac{ \omega+1}{\omega+2} =\frac{ (A+B)^2}{2(1+AB)} \,,\label{stocz}
\ee
where $A,B$, and $\eta$ are real constants. In~\cite{ALC}, this is assumed 
to be a gauge choice valid for {\em 
any} solution satisfying~1)-3), but at this stage this is instead  
a choice of a special 
solution, the Campanelli-Lousto one.  The general form of 
the Campanelli-Lousto solution of Eqs.~(\ref{field1}) and (\ref{field2}) 
is \cite{CampanelliLousto} 
\begin{eqnarray}
ds^2_{CL} &=& -\left( 1-\frac{2\eta}{r}\right)^{b_0+1}dt^2 + 
\left( 1-\frac{2\eta}{r}\right)^{-a_0-1} dr^2  \nonumber\\
&&\nonumber\\
&\, & + \left( 1-\frac{2\eta}{r}\right)^{-a_0}r^2 d\Omega_{(2)}^2 \,, 
\label{CL1}\\
&&\nonumber\\
\phi_{CL}(r) &=& \phi_0 \left( 
1-\frac{2\eta}{r}\right)^{\frac{a_0-b_0}{2}} \,,\label{CL2}
\end{eqnarray}
where $a_0$ and $b_0$ are two parameters satisfying 
\be
\omega =\frac{-2\left(a_0^2 +b_0^2 -a_0b_0 
+a_0+b_0\right)}{\left(a_0-b_0 \right)^2} \,. \label{omegaab}
\ee
It is clear that the Agnese-La Camera choice is reproduced for 
$ a_0 = -B-1 $ and $ b_0 = A-1$. Therefore, the results of \cite{ALC} are 
true only for this 
particular solution (in spite 
of being advertised as black holes, the Campanelli-Lousto family contains 
only wormhole throats and naked singularities \cite{Vanzo}). The 
conflict with  
the no-hair theorems is then resolved. But what are the 
solutions satisfying~1)-3) which 
are not Schwarzschild?

\subsection{The general solution}

Let $\left( g_{ab}, \phi \right)$ be a solution under the 
assumptions~1)-3). By performing the standard conformal transformation to 
the 
Einstein frame representation of Brans-Dicke gravity
\begin{eqnarray}
g_{ab}& \rightarrow& \tilde{g}_{ab}=\Omega^2 \, g_{ab}= \phi g_{ab} \,, 
\label{CT1}\\
&&\nonumber\\
\phi & \rightarrow& \tilde{\phi} =\sqrt{ \frac{|2\omega+3|}{16\pi}} \, 
\ln\left( \frac{\phi}{\phi_0} \right) \label{CT2}
\end{eqnarray}
(where $\phi_0$ is a constant), the Brans-Dicke action~(\ref{action}) is 
recast in the form
\be
S_{BD}= \int d^4 x \sqrt{-\tilde{g}} \left( \frac{ \tilde{\cal{R}}}{16\pi} 
-\frac{1}{2} \, \tilde{g}^{ab} \tilde{\nabla}_a \tilde{\phi} 
\tilde{\nabla}_b \tilde{\phi} \right) \,.\label{Eframe-action}
\ee
Since the conformal factor is $\Omega=\sqrt{\phi(r)}$, the 
Einstein frame geometry is also spherical, static, and asymptotically 
flat. Formally, the 
action~(\ref{Eframe-action}) describes GR 
with a free, minimally coupled scalar field and the  most 
general spherical, static, asymptotically flat solution is known to be the 
Fisher-Janis-Newman-Winicour-Buchdahl-Wyman (FJNWBW) solution of 
GR \cite{Fisher,Wyman}   
\begin{eqnarray}\label{FJNWBW}
d\tilde{s}^2 &=& -\mbox{e}^{ \alpha/r} dt^2 + \mbox{e}^{-\alpha/r}
\left( \frac{ \gamma/r}{\sinh( \gamma/r)}\right)^4 dr^2 \nonumber\\
&&\nonumber\\
&\, &  +  \mbox{e}^{-\alpha/r} \left( \frac{ \gamma/r}{\sinh( 
\gamma/r)}\right)^2   r^2 d\Omega_{(2)}^2
\end{eqnarray}
(where $\alpha $ and $\gamma$ are constants) with scalar field 
\cite{Fisher,Wyman} 
\be\label{F1}
\tilde{\phi} = \frac{\phi_*}{r} \,, \;\;\;\;\;\;\;
\phi_*=\frac{-\sigma}{4\sqrt{\pi}} \,,
\ee
where $\sigma$ is a scalar charge and one can take $\gamma \geq 0$ 
without loss of generality. These three constants are related 
by \cite{Wyman}   
\be\label{WymanRelation}
4\gamma^2=\alpha^2+2\sigma^2 \,.
\ee
If $\sigma=0$, the 
Einstein frame scalar vanishes, the Jordan frame scalar reduces to a 
constant, the theory reduces to GR, and the solution 
reduces to Minkowski   
in both conformal frames, which then coincide. In fact, the constants 
$\alpha$ and $\gamma$ both vanish whenever $\sigma$ does, thus turning   
(\ref{FJNWBW}) into the Minkowski metric. However, as the 
notation followed here is that of Ref.~\cite{Wyman},  the relation 
(\ref{WymanRelation}) between the constants $\alpha$, $\gamma$ and 
$\sigma$ does not allow one 
to see this fact as it only implies $4\gamma^2=\alpha^2$ when $\sigma=0$.  
The vanishing of $\alpha$ and $\gamma$ could be seen only when tracing 
back the steps that led to expression (\ref{FJNWBW}) as presented in 
Ref.~\cite{Wyman}, for then one clearly sees that whenever $\sigma$  
vanishes, so does the constant 
$\alpha$, which, in turn, makes $\gamma$ vanish as well.

Consider now 
the case 
$\sigma \neq 0$. 
Mapping the FJNWBW solution back to the Jordan frame, one obtains the 
most general 
solution of the Brans-Dicke equations under the 
assumptions~1)-3) (a remark to this regard was made in 
passing in \cite{BhadraSarkar}). Equation~(\ref{CT2}) yields the scalar 
field \be
\phi (r) = \phi_0 \, \mbox{e}^{-\beta/r}  \,,
\;\;\;\;\;\;\; \beta= \frac{\sigma}{\sqrt{|2\omega+3|} }  \,,
\label{new2}
\ee
while Eq.~(\ref{CT1}) gives 
\begin{eqnarray}
ds^2 &=&  -\mbox{e}^{ (\alpha+\beta)/r } dt^2
+ \mbox{e}^{ ( \beta-\alpha)/r  }
\left( \frac{ \gamma/r }{ \sinh ( \gamma/r ) } \right)^4 dr^2  \nonumber\\
&&\nonumber\\
&\, & + \mbox{e}^{ (\beta-\alpha)/r }  
 \left(
\frac{ \gamma/r }{ \sinh( \gamma/r) } \right)^2 r^2 d\Omega_{(2)}^2 
\,. \label{new1}
\end{eqnarray}
This is the most general solution of Brans-Dicke theory under the 
assumptions~1)-3). It is related to a Campanelli-Lousto solution. The 
special case $\gamma=0$ will be discussed later.

It should be noted here that in Ref.~\cite{Bronnikov2}, the most 
general solution of the generalized Brans-Dicke scalar-tensor theory has 
also been found by Bronnikov for the case of electrovacuum. For the case  
of vacuum, explicit forms of the solution corresponding to  
imaginary $\gamma$,    for 
which the $\sinh$ function in the metric (\ref{new1}) is replaced 
by the sine function, were given there. The latter possibility, 
corresponding to 
what has been called in 
Refs.~\cite{ColdBH} a ``cold black hole'', 
arises for the anomalous case, $2\omega+3<0$. This case, which we 
avoided in this paper by taking care of using the absolute value of 
$2\omega+3$ in our field redefinition (\ref{CT2}), is anomalous for it 
makes the Einstein frame field $\tilde{\phi}$ imaginary which, in turn, 
makes the kinetic term in the Einstein frame action  
(\ref{Eframe-action}) acquire the 
wrong sign. This case gives the ghost counterpart of the 
solution~(\ref{FJNWBW}) and (\ref{F1}) due to Bergman and  Leipnik 
\cite{Fisher}.   
Indeed, when an imaginary field 
$\tilde{\phi}$ is allowed, 
the scalar charge $\sigma$ becomes imaginary and the Wyman relation 
between the various constants becomes \cite{Bronnikov2} 
\be
-4\gamma^2=\alpha^2-2\sigma^2 \,.
\ee
The negative signs can be absorbed by 
letting both $\sigma$ and $\gamma$ be imaginary.  This then turns the 
$\sinh$ function into a sine function in (\ref{new1}).\footnote{For 
completeness, we give here the general solution of Brans-Dicke theory in 
the anomalous case. It reads,\\ $ds^2=-\mbox{e}^{\frac{\alpha+\beta}{r}} dt^2
+ \mbox{e}^{\frac{\beta-\alpha}{r}}
\!\!\left(\frac{\gamma/r}{\sin(\gamma/r)} 
\right)^2\left[\left(\frac{\gamma/r}{\sin(\gamma/r)}\right)^2dr^2  + r^2 
d\Omega_{(2)}^2\right]$.} In Ref.~\cite{Bergh}, the 
special cases $\alpha=\beta$, $\alpha=(2\omega+3)\beta$, and 
$\alpha=-(\omega+1)\beta$ in (\ref{new1}) were  found explicitly. Much 
later, a more exhaustive investigation of the general solutions of the 
Bergmann-Wagoner class of scalar-tensor theories, in which Brans-Dicke 
gravity is a special case, was made in Ref.~\cite{Bronnikov3}. It was 
shown there that among these solutions black hole geometries arise for the 
anomalous versions of these theories. The thermodynamics of such black 
holes, also dubbed ``cold black holes'', were investigated. 

Let us now come back to our general solution. When 
$\gamma\neq 0$ in (\ref{new1}), by performing the two consecutive coordinate 
transformations 
\begin{equation} 
\label{CoordTransformations}
\mbox{e}^{\gamma/r} = \frac{1+B/\rho}{1-B/\rho} \,,\qquad
\bar{r} = \rho \left( 1+\frac{B}{\rho} \right)^2 \,,
\end{equation}
and setting $\eta=2B=\sqrt{m^2+\sigma^2} \,,  m/\eta =-\alpha/(2\gamma)$, 
$ \sigma/\eta=\beta \sqrt{|2\omega+3|}/(2\gamma)$,  and rescaling the 
time coordinate by a factor $|\gamma/(2B)|$, the 
solution~(\ref{new1}), (\ref{new2}) becomes 
\begin{eqnarray}
ds^2 &=& -\left( 1-\frac{2\eta}{\bar{r}} \right)^{ \frac{1}{\eta} \left( 
m-\frac{\sigma}{ \sqrt{|2\omega+3|}}\right)} dt^2 \nonumber\\
&&\nonumber\\
&\, & + \left( 1-\frac{2\eta}{\bar{r}} \right)^{ \frac{-1}{\eta} \left(
m+\frac{ \sigma}{ \sqrt{|2\omega+3|} } \right)} d\bar{r}^2 \nonumber\\
&&\nonumber\\
&\, & + \left(
1-\frac{2\eta}{\bar{r}} \right)^{1 - \, \frac{1}{\eta} \left(
m+\frac{\sigma}{ \sqrt{|2\omega+3|}}\right)} \bar{r}^2 d\Omega_{(2)}^2 
\,, \label{CL2_1}\\
&&\nonumber\\
\phi & = & \phi_0 \left(
1-\frac{2\eta}{\bar{r}} \right)^{ \frac{\sigma}{\eta \sqrt{ |2\omega+3|} } 
} \,,\label{CL2_2}
\end{eqnarray}
which is a Campanelli-Lousto solution with 
\begin{eqnarray}
a_0 &=& -1+\frac{1}{\eta} \left( m+\frac{\sigma}{ 
\sqrt{|2\omega+3|}}\right) \,,\\
&&\nonumber\\ 
b_0 &=&  -1+\frac{1}{\eta} \left( 
m-\frac{\sigma}{\sqrt{|2\omega+3|}}\right) \,.
\end{eqnarray}
It must be noted here that although this form contains only the absolute 
value of the term $2\omega+3$, the anomalous case $2\omega+3<0$ discussed 
above does not apply here as the coordinate redefinitions 
(\ref{CoordTransformations}) would not be real-valued anymore since 
$\gamma$ is imaginary in this case. Therefore, the Campanelli-Lousto 
metric (\ref{CL1}), as well as its other version (\ref{ALC1}) used in 
Ref.~\cite{ALC}, are only valid for the normal case $2\omega+3>0$.

\subsection{Generality of the solution}\label{Generality}

It seems that, given a solution $\left( g_{ab}, \phi \right)$ of the form 
(\ref{CL2_1}), (\ref{CL2_2}), one could still change the scalar 
field according to $\phi \rightarrow \Phi =\phi \, 
\mbox{e}^{\psi(r)}$ in such a way that $\left( g_{ab}, \Phi 
\right)$ is still a solution, which would mean that (\ref{new1}), 
(\ref{new2}) do not give all the 
possible solutions, and hence do not constitute the most general 
solution. We show here that  this is not the case. In fact, the Ricci 
tensor component
\be
R_{rr}=\omega \left( \frac{\nabla_r \phi }{\phi} \right)^2 +\frac{ 
\nabla_r \nabla_r \phi}{\phi} 
\ee 
does not change when $\left( g_{ab}, \phi \right)$ is changed into 
$\left( g_{ab}, \Phi  \right)$ provided that 
\be
\left(\omega+1\right) \left( \nabla_r \psi\right)^2 +2(\omega+1) 
\frac{\nabla_r \phi }{\phi} \, 
\nabla_r \psi +\nabla_r \nabla_r \psi =0 \,, \label{czz1}
\ee
while it must be $
\Box \left( \phi \, \mbox{e}^{\psi} \right)=0   $ 
in order for $\Phi$ to still be a solution.  Using this equation 
to eliminate $\nabla_r \nabla_r\psi$ in Eq.~(\ref{czz1}), one finds 
\be
 \left( \nabla_r \psi + \frac{2\nabla_r \phi}{ \phi} 
\right) \nabla_r \psi  =0 \,,
\ee
which (apart from the irrelevant possibility  $\psi=$~const.)  
integrates to $\psi(r)=-2\ln\phi +\mbox{const.}$ and $\Phi=C_0/\phi$. 
However,  replacing $\left( g_{ab}, \phi  
\right)$ with $\left( g_{ab}, C_0/\phi \right)$ amounts to changing 
the exponent $\beta$ in Eq.~(\ref{new2}) into $-\beta$, or to changing 
the sign of the  scalar charge $\sigma$,  a possibility already included 
in the form of the general solution (\ref{new1}), (\ref{new2}). 

It must be noted, however, that what makes the conformal 
transformation one to one and protects (\ref{new1}) and (\ref{new2}) 
against such redefinitions as $\phi\longrightarrow\phi\, e^{\psi(r)}$, 
that could have prevented them from being the most general solution, is the 
homogeneous wave equation $\Box\phi=0$. The latter, in turn, is always 
guaranteed to hold in vacuo and electrovacuo for  which 
the matter energy-momentum tensor is traceless. Therefore, we conclude 
that (\ref{new1}) and (\ref{new2}) constitute indeed the most general 
solution with the assumptions~1)-3) above. Moreover, this analysis also 
applies, and therefore reinforces, Bronnikov's general solutions  for 
vacuum and electrovacuum scalar-tensor theories found in 
Ref.~\cite{Bronnikov2}.

\subsection{Nature of the solution} 

In this subsection we  investigate the nature of the general solution 
(\ref{new1}) and (\ref{new2}).  To assess whether the general 
geometry~(\ref{new1}) describes black holes, 
wormholes, or naked  singularities, one examines the horizons (if they 
exist) and their nature.  The equation we are going to use for locating 
the apparent horizons is 
\cite{MisnerSharp, Visser} $\nabla^c R \nabla_c R=0 $,
where 
\be
R(r)= \gamma\, \frac{ \mbox{e}^{\frac{\beta-\alpha}{2r}} 
}{\sinh(\gamma/r)}  \label{arealradius}
\ee
is the areal radius. Horizons correspond to the roots 
of that equation; a 
single root describes a black hole horizon while a double root describes a 
wormhole throat.  With (\ref{arealradius}), the equation becomes 
\be 
g^{rr} \left( \frac{dR}{dr} \right)^2 = 
\sinh^2(\gamma/r)  \left[  \frac{\alpha -\beta}{2\gamma} + \coth 
\left( \gamma/ r \right)   \right]^2  =0 \,. 
\ee
It is clear that, if roots exist, they are always double roots 
corresponding to wormhole throats. They exist if $(\beta-\alpha)/\gamma 
>0$ and, in this case, they are given by
\be
r_H= \frac{2\gamma}{\ln \left( 
\frac{\beta-\alpha+2\gamma}{\beta-\alpha-2\gamma}\right)} 
= \frac{\gamma}{ \tanh^{-1}\left( \frac{2\gamma}{\beta-\alpha}\right)}
\,, 
\ee
If $(\beta -\alpha)/\gamma<0$, instead, there is a naked singularity. 
In fact, the general solution~(\ref{new1}) has a spacetime singularity at 
$R=0$, as is deduced from the Ricci scalar 
\be
{\cal R}= \frac{\omega}{\phi^2} \nabla^c \phi \nabla_c\phi 
=\left\{ 
\begin{array}{cc}
 \frac{ \omega \beta^2}{\gamma^4}  \, \mbox{e}^{\left( 
\alpha-\beta\right)/ r }  
\sinh^4(\gamma/r) & \;\; \mbox{if} \; \gamma\neq 0  \,,\\
&\\
\frac{ \omega \beta^2}{r^4}  \, \mbox{e}^{\left( 
\alpha-\beta\right)/ r }  & \;\; \mbox{if} \; \gamma= 0  \,,
\end{array} \right.
\ee
If $\gamma\neq 0$ then when $r\rightarrow 0$ we have, depending 
on whether $\gamma $ is positive or negative,
\be
{\cal R}= \frac{\omega \beta^2}{16\gamma^4} \, \mbox{e}^{\left( 
\alpha-\beta \pm 4\gamma\right)/r} \,,
\ee
respectively. Therefore, the Ricci scalar diverges as $r \rightarrow 0$ 
only for $\beta -\alpha<4\gamma$ or for $\alpha-\beta > 4\gamma$, 
respectively.

In the special case $\gamma=0$, the FJNWBW metric reduces to the Yilmaz 
geometry \cite{Yilmaz} and its Jordan  frame  
cousin~(\ref{new1}), (\ref{new2}) is the Brans Class~IV solution 
\cite{Brans} 
\begin{eqnarray}
ds^2 &=& -\mbox{e}^{ -2B/r }dt^2 + \mbox{e}^{ 2B(C+1)/r }
\left( dr^2+r^2 d\Omega_{(2)}^2 \right),\\
&&\nonumber\\
\phi&=& \phi_0 \, \mbox{e}^{-BC/r} \,,
\end{eqnarray}
where $B=-(\alpha+\beta)/2, C=-2\beta/(\alpha+\beta)$. The equation locating the apparent horizons reduces to $ 
\left( 1- \frac{\beta-\alpha}{2r} \right)^2=0 $, 
which has  a double root $r_H=(\beta-\alpha)/2$ corresponding to a  
wormhole throat if $\beta>\alpha$ and to a central naked singularity 
otherwise.

These results about the nature of the solutions of 
Brans-Dicke theory have already been worked out in detail in \cite{VFS}. 
Therefore, this analysis satisfactorily shows that it is possible to use 
the most general solution (\ref{new1}) of Brans-Dicke theory to recover 
in a compact way the results already found in Ref.~\cite{VFS} by 
going through each of 
the Brans classes of solutions individually. The general solution 
(\ref{new1}) 
thus allows for a unified investigation of the physics behind the 
four Brans classes of solutions.

It must be noted here that, just as it was done in Ref.~\cite{VFS}, the 
investigation of the nature of the solution conducted here is 
based on the simple detection of possible black hole horizons or wormhole 
throats. In fact, in contrast to the analysis made in 
Ref.~\cite{Wormholes}, no additional requirements, such as asymptotic 
flatness or regularity of the spacetime away from the throat when the 
latter exists, are imposed before calling such a solution a wormhole. The 
wormhole definition that is implicitly adopted here, and which was also 
adopted in Ref.~\cite{VFS}, is that of Ref.~\cite{HochbergVisser} which 
consists of a quasi-local definition involving only the properties of the 
local geometry of spacetime. Of course, ``extended'' wormholes can 
also be studied, and they are related to the ``cold black holes'' of 
\cite{ColdBH}. We refer the reader to \cite{ColdBH} for these 
situations.

\section{Conclusions} 
\label{sec:4}

The key to solving the Brans-Dicke equations under the assumptions~1)-3) 
is to map the problem into the Einstein frame and using a known result of 
Einstein-massless Klein-Gordon theory \cite{Fisher}. By contrast, little 
progress is made when analyzing directly the Jordan frame field equations.  
The previous section shows that the Schwarzschild black hole is obtained 
when the scalar charge $\sigma$ vanishes and that there is no other black 
hole solution under the assumptions made. This result matches the no-hair 
theorems of \cite{Hawking, Bekenstein, SotiriouFaraoniPRL, 
BhattaRomanoPRL, JFnohair} (which are, however, more general). The 
remaining solutions, corresponding to $\sigma\neq 0$, are necessarily of 
the Campanelli-Lousto form~(\ref{CL1}), (\ref{CL2}) or conformal to it. 
They can only 
describe 
wormhole throats or naked singularities, according to the values of 
the parameters $\sigma/m$ and $\omega$ (or of $\alpha$ and $\beta$). 

The most general 
solution of Jordan frame 
Brans-Dicke theory under the assumptions~1)-3) is given by 
Eqs.~(\ref{new1}) and (\ref{new2}) and is conformal to the FJNWBW 
solution 
of GR. Our analysis in Sec.~\ref{Generality} established the 
general character of this solution and pointed to the homogeneous wave 
equation that the scalar field obeys as being the ingredient that  renders 
the solution  (\ref{new1}) really general. We also pointed out that it is 
only thanks to this constraint that the conformal transformation trick 
remains a one-to-one mapping and allows one to extract the general 
Jordan frame solution from the most general Einstein 
frame one. The homogeneity of the wave equation, being guaranteed by the 
tracelessness of the matter energy-momentum tensor, makes the conformal 
trick work both in vacuum and electrovacuum. The conformal 
transformation trick is indeed what has allowed Bronnikov in 
Ref.~\cite{Bronnikov2} to extract the general solution for the 
electrovacuum case as well. 

We have investigated the general solution of vacuum Brans-Dicke  
gravity, a result that 
constitutes an analogue of the Jebsen-Birkhoff theorem of GR in 
Brans-Dicke theory, which is a rarity in alternative gravities. In 
principle, this result can be 
circumvented in the same ways already conceived to evade the no-hair 
theorems for scalar-tensor black holes: by 
including matter, by allowing the scalar field to depend on time while 
keeping the geometry static, or by letting the scalar field diverge or 
vanish on the horizons \cite{circumvent}. 

As already noted above,  generalizations of the results 
 presented here have already been obtained. The generalization of 
the Jebsen-Birkhoff theorem to multidimensional GR was given in 
Ref.~\cite{Bronnikov1}, and the general spherically symmetric solution of 
the Bergmann-Wagoner class of scalar-tensor theories, of which Brans-Dicke 
gravity is a special case, has been found in Ref.~\cite{Bronnikov3}. 
Generalizations to situations in which a cosmological constant or 
a 
non-gravitational scalar field are present and, more important, to axial 
symmetry, may be possible and will be explored elsewhere.

{\em Acknowledgments:} The authors are grateful to K.A. Bronnikov for 
bringing to their attention Refs.~\cite{Bronnikov1,Bronnikov2,Bronnikov3}, 
to N. Van Den Bergh for communicating Ref.~\cite{Bergh}, and to an 
anonymous referee for useful comments and references. This work is 
supported by the Natural Sciences and Engineering Research Council of 
Canada (Grants No.~2016-03803 and No.~2017-05388).

\end{document}